\documentclass[10pt]{article}

\usepackage{bm,graphicx,amsmath,amssymb,url}

\begin{document}

\title{On the Kohn--Sham density response in a localized basis set}
\author{Dietrich Foerster\footnote{d.foerster@cpmoh.u-bordeaux1.fr},
Peter Koval\footnote{koval.peter@gmail.com}
\footnote{\textit{CPMOH, University of Bordeaux 1, 351 Cours de la Liberation, 33405, Talence, France}}}

\maketitle

\begin{abstract}
We construct the Kohn--Sham density response function $\chi_{0}$ in a
previously described basis of the space of orbital products. The
calculational complexity of our construction is $O(N^{2}N_{\omega})$ for a
molecule of $N$ atoms and in a spectroscopic window of $N_{\omega}$
frequency points. As a first application, we use $\chi_{0}$ to calculate molecular
spectra from the Petersilka--Gossmann--Gross equation. With $\chi_{0}$ as
input, we obtain correct spectra with an extra computational effort that
grows also as $O(N^2 N_{\omega})$ and, therefore, less steeply in $N$ than
the $O(N^{3})$ complexity of solving Casida's equations. Our construction
should be useful for the study of excitons in molecular physics and in
related areas where $\chi_{0}$ is a crucial ingredient.
\newline
\textbf{Published in:} D.~Foerster and P.~Koval, J.~Chem.~Phys. \textbf{131}, 044103 (2009).\newline
\url{http://link.aip.org/link/?JCP/131/044103}\newline
\texttt{DOI: 10.1063/1.3179755}
\newline
\texttt{PACS: 33.20.-t, 31.15.ee}
\newline
\texttt{Kohn--Sham density response, Petersilka--Gossmann--Gross equation,
iterative Lanczos method}
\end{abstract}

\section{Introduction and motivation}

A basic concept in time-dependent density functional theory \cite{DFT,TDDFT}
is a reference system of noninteracting electrons of the same density
$n(\bm{r},t)$ as the interacting electrons under study and which move in an
appropriately adjusted potential $V_{\mathrm{KS}}(\bm{r},t)$. Therefore, an
important element of this theory is the density response function $\chi_{0}(%
\bm{r},\bm{r}^{\prime},t-t^{\prime})$ that describes the variation in
density $\delta n(\bm{r},t)$ of such reference electrons upon a change $%
\delta V_{\mathrm{KS}}(\bm{r}^{\prime},t^{\prime})$ of the Kohn--Sham
potential 
\begin{equation}
\chi_{0}(\bm{r},\bm{r}^{\prime},t-t^{\prime})=\frac{\delta n(\bm{r},t)}{%
\delta V_{\mathrm{KS}}(\bm{r}^{\prime},t^{\prime})}.
\label{density_response_definition}
\end{equation}
The Kohn--Sham density response is needed for implementing Hedin's GW
approximation \cite{Hedin:65}, for electronic excitation spectra \cite%
{GrossPetersilka,Casida}, for treating excitons in molecular systems
\cite{Tiago}, and in other contexts, such as the inclusion of the van der Waals
interaction in DFT \cite{vanderWaals}. Various methods have been developed
for constructing this response function in solids \cite{LuciaMustBeQuoted},
but for molecules no computationally efficient method has emerged.
Therefore, the Kohn--Sham density response remains an important bottleneck
in applications of electronic structure methods to molecular physics. In the
present paper we describe a solution to this long standing technical problem.

The difficulty of dealing with the noninteracting density response is
somewhat surprising, since it can be written down very compactly in terms of
molecular orbitals \footnote{The Fourier transform of this time ordered
expression agrees with that of the retarded correlator at positive frequencies.}

\begin{equation}
\chi_{0}(\bm{r},\bm{r}^{\prime},\omega)= \sum_{E,F;E\cdot F<0}\frac{n_{F}-n_{E}}{%
\omega-(E-F)-\mathrm{i}\varepsilon (n_{E}-n_{F})} \varphi^{E}(\bm{r})
\varphi^{F}(\bm{r}) \varphi^{E}(\bm{r}^{\prime})\varphi^{F}(\bm{r}^{\prime}).
\label{conventional}
\end{equation}%
Here $\varphi^{E}(\bm{r})$ represents a stationary (real valued) molecular
orbital of energy $E$ measured relative to the Fermi energy (set to zero), $n_{E}, n_{F}$
are occupation factors and the energies $E,F$ of a particle--hole pair must be
of opposite signs, $E\cdot F<0$. 
Although this form of the density response was used very effectively in Casida's equations
for molecular spectra \cite{Casida}, it is less useful, for example, in
computing the screening of the Coulomb interaction. In this context, we must
integrate over the arguments $\bm{r}$, $\bm{r}^{\prime}$ of $\chi_{0}(\bm{r},%
\bm{r},\omega^{\prime})$ which requires a summation over $O(N^{2})$ pairs of
points $\bm{r}$, $\bm{r}^{\prime}$ and over $O(N^2)$ energies $(E,F)$ where $%
N$ is the number of atoms. Therefore, a straightforward application of the
conventional expression requires a total of $O(N^4 N_{\omega})$ operations.
The strong growth of CPU (central processing unit) effort with the number of atoms limits the
usefulness of expression (\ref{conventional}) to molecules or clusters
containing very few atoms.

Eq.~(\ref{conventional}) shows that $\chi_{0}$ acts in the space of products
of molecular orbitals $\varphi^{E}(\bm{r}) \varphi^{F}(\bm{r})$, a space
that has no obvious basis. Chemists have long recognized that both the space
of products of molecular orbitals and the related space of products of
atomic orbitals contain many linearly dependent elements \cite{Harriman}. To
eliminate such redundant elements, products of orbitals are usually
parametrized in terms of sets of auxiliary functions \cite{Boeij:2008},\cite{Casida}.

In a better controlled and more systematic approach \cite{DF}
(for a similar method in the context of the GW approach see
\cite{Aryasetiawan-and-Gunnarsson:1994}), one identifies
the dominant elements in the space of all products of a given pair of atoms.
As a result of this construction, any product of atomic orbitals
$f^{a}(\bm{r})$, $f^{b}(\bm{r})$ can be expanded in a basis of $O(N)$
``dominant functions'' $\{F^{\mu}(\bm{r})\}$. 
In this basis, the density
response acts as a frequency dependent matrix $\chi^0_{\mu\nu}(\omega)$ 
\begin{equation}
\chi_{0}(\bm{r},\bm{r}^{\prime},\omega)=\sum_{\mu,\nu }F^{\mu}(\bm{r})
\chi^0_{\mu\nu}(\omega)F^{\nu}(\bm{r}^{\prime}).  \label{TensorForm}
\end{equation}
The present paper describes an efficient construction of this matrix by
Green's function type methods that require $O(N^{2}N_{\omega })$ operations
for a molecule of $N$ atoms and in a spectroscopic window of $N_{\omega}$
frequency points.

To test this construction, we apply it to electronic excitation spectra of
molecules, where many results are known. We consider two approaches for
excitation spectra: the Petersilka--Gossmann--Gross equations
\cite{GrossPetersilka} and Casida's equations \cite{Casida}.
The test of $\chi_{0}$ on molecular spectra turns out to be successful
as our spectra agree indeed with those found from Casida's equation.

Our paper is organized as follows: in section \ref{s:2} we derive a spectral
representation of $\chi_{0}$ and discuss its locality properties.
In section \ref{s:3}, we formulate and test an algorithm that exploits
the spectral representation of $\chi_{0}$.
In section \ref{s:4}, we further test $\chi_{0}$ by applying it to the
computation of electronic excitation spectra. To accelerate the computation,
we develop an iterative Lanczos-like procedure.
Section \ref{s:5} gives our conclusions.

\section{A spectral representation for the Kohn--Sham density response}

\label{s:2}

\subsubsection*{Extended versus local fermions}

Our approach is based on Green's functions and their spectral functions. So
let us recall some of their basic definitions \cite{Fetter-Walecka:1971} and
establish our notation \footnote{%
Atomic units are used throughout in this paper.}. In the framework of second
quantization, the electronic propagator reads

\begin{equation}
\mathrm{i}G(\bm{r},\bm{r}^{\prime},t-t^{\prime})=\left\langle 0\left|
T\left\{\psi(\bm{r},t)\psi^{+}(\bm{r}^{\prime},t^{\prime})\right\}
\right|0\right\rangle= \theta(t -t^{\prime})\langle 0|\psi(\bm{r}%
,t)\psi^{+}(\bm{r}^{\prime},t^{\prime})|0\rangle - \theta(t^{\prime}-t
)\langle 0|\psi^{+}(\bm{r}^{\prime},t^{\prime})\psi(\bm{r},t)|0\rangle,
\label{Green_Definition}
\end{equation}
where $\psi$ and $\psi^{+}$ are annihilation and creation operators of
electrons, respectively. The symbol $T$ represents time ordering of
operators and $\theta(t)$ is the unit step function.

According to general principles \cite{Fetter-Walecka:1971}, the density
response function (\ref{density_response_definition}) coincides with the
density--density correlator of the unperturbed system 
\begin{equation}
\mathrm{i}{\chi_{0}}(\bm{r},\bm{r}^{\prime},t-t^{\prime}) =\langle 0| T\{ n(%
\bm{r},t) n(\bm{r}^{\prime},t^{\prime})\}|0\rangle,
\label{chi_product_green}
\end{equation}
where $n=\psi^{+}\psi$ is the electronic density operator. For simplicity,
we mostly use the time ordered form of correlators, the Fourier transform of
which coincides with that of the causal one at positive frequencies. For
noninteracting electrons, the density response can be expressed in terms of
electron propagators $G(\bm{r},\bm{r}^{\prime},t-t^{\prime})$. Applying
Wick's theorem \cite{Fetter-Walecka:1971} on eq.~(\ref{chi_product_green}),
we find

\begin{equation}
\mathrm{i}{\chi_{0}}(\bm{r},\bm{r}^{\prime},t-t^{\prime})=G(\bm{r}^{\prime},%
\bm{r},t-t^{\prime})G(\bm{r}^{\prime},\bm{r},t^{\prime}-t),
\label{chi0-gf-gf}
\end{equation}
where we ignore the time-independent (disconnected) part of the correlator
that contributes to the response only at zero frequency.

To confirm the conventional expression for ${\chi_{0}}$ in terms of
molecular orbitals (\ref{conventional}), we expand the operators $\psi(\bm{r}%
,t)$ in terms of Kohn--Sham orbitals $\varphi_E(\bm{r})$ and their
associated fermion operators $c_E(t)$ 
\begin{equation}
\psi(\bm{r},t)=\sum_{E}\varphi_E(\bm{r}) c_{E}(t).  \label{psi-sum}
\end{equation}
We use the last expression to rewrite the Green's function (\ref%
{Green_Definition}) in terms of molecular orbitals 
\begin{equation}
G(\bm{r},\bm{r}^{\prime},t-t^{\prime})=-\mathrm{i}\theta
(t-t^{\prime})\sum_{E>0}\varphi_{E}(\bm{r}) \varphi_{E}(\bm{r}^{\prime})%
\mathrm{e}^{-\mathrm{i}E(t-t^{\prime})}+\mathrm{i}\theta (t^{\prime
}-t)\sum_{E<0}\varphi_{E}(\bm{r})\varphi_{E}(\bm{r}^{\prime})\mathrm{e}^{-%
\mathrm{i}E(t-t^{\prime})},  \label{traditional_Green}
\end{equation}%
where we took into account the anticommutator $\left[ c_{E}(t),c_{E^{%
\prime}}^{+}(t)\right]_{+}=\delta_{E,E^{\prime}}$, the time evolution $%
c_{E}(t)=\mathrm{e}^{-iEt}c_{E}(0)$ and the nature of the ground state.
Regularizing the above expression with a damping factor $\mathrm{e}%
^{-\varepsilon(t-t^{\prime})/2}$, using (\ref{traditional_Green}) in eq. (%
\ref{chi0-gf-gf}) and doing a Fourier transform on the result, we easily
confirm the textbook expression eq.~(\ref{conventional}) for ${\chi_{0}}$.

Our approach emphasizes locality and it is better to use a localized basis
of atomic orbitals $f^{a}(\bm{r})$. Therefore, we write the molecular
orbitals as linear combinations of atomic orbitals (LCAO) 
\begin{equation}
\varphi_{E}(\bm{r})=\sum_{a}X_{a}^{E}f^{a}(\bm{r}).  \label{lcao}
\end{equation}
Here $X_{a}^{E}$ are (generalized) eigenvectors of the Kohn-Sham Hamiltonian
labelled by their eigenvalues $E$. Inserting the latter expression in
equation (\ref{traditional_Green}), we obtain the propagator in the
localized basis. For later convenience, we write this result in terms of
spectral functions of particles and holes 
\begin{equation}
G_{ab}(t) =-\mathrm{i}\theta (t)\int_{0}^{\infty }\,ds\ \rho_{ab}^{+}(s)%
\mathrm{e}^{-\mathrm{i}st}+ \mathrm{i}\theta (-t)\int_{-\infty }^{0}\,ds\
\rho_{ab}^{-}(s)\mathrm{e}^{-\mathrm{i}st}.  \label{gf}
\end{equation}
Here, we introduced the spectral densities of particles and holes 
\begin{equation}
\rho_{ab}^{+}(s)=\sum_{E>0} X_{a}^{E} X_{b}^{E}\delta(s-E)\text{ \ and }%
\rho_{ab}^{-}(s)=\sum_{E<0} X_{a}^{E} X_{b}^{E}\delta(s-E).  \label{sf}
\end{equation}

\subsubsection*{Definition of a frequency dependent response matrix}

Let us also write the operators $\psi(\bm{r},t)$ in terms of localized
atomic orbitals with localized fermion operators $c_a(t)$ as coefficients 
\cite{Fulde}. We have 
\begin{equation}
\psi(\bm{r},t)=\sum_{a} f^{a}(r) c_{a}(t)  \label{LocalFermions-1}
\end{equation}
with $c_{a}(t)=\sum_{E}X_{a}^{E}c_{E}(t)$. Because we use local orbital
fermions (\ref{LocalFermions-1}), the electron density $n(\bm{r},t)$ is
given by a sum over products of orbitals multiplied by bilinears of fermion
operators 
\begin{equation}
n(\bm{r},t)=\psi^{+}(\bm{r},t)\psi(\bm{r},t)= \sum_{a,b}f^{a}(\bm{r})f^{b}(%
\bm{r})c_{a}^{+}(t) c_{b}(t).  \label{bilinear}
\end{equation}%
It is well known that the set of products of orbitals contains collinear or
nearly collinear elements, a fact nicely illustrated by taking products of
the eigenfunctions of the harmonic oscillator \cite{Harriman}.
Traditionally, this difficulty is treated by expanding products
of orbitals in auxiliary functions.

In a recent and more systematic approach \cite{DF}, one identifies a set of
``dominant products'' $\{F^{\lambda }(\bm{r})\}$ as special linear
combinations in the space of products of orbitals. 
A similar method was previously developed in the context of the GW method 
\cite{Aryasetiawan-and-Gunnarsson:1994}. The main collinearity of the set of
orbital products occurs at the level of a fixed pair of atoms. Therefore all
the products $f^{a}(\bm{r})\cdot f^{b}(\bm{r})$ belonging to such a fixed
pair were formed. A matrix of overlaps between these products was computed
and the dominant products were found among the linear combinations of the
original products that diagonalize this matrix.
As a result, any nonzero product of orbitals belonging to a given pair of
atoms can be expressed in terms of a much smaller set of dominant products
with respect to the same pair of atoms
\begin{equation}
f^{a}(\bm{r})f^{b}(\bm{r})=\sum_{\lambda }V_{\lambda }^{ab}F^{\lambda }(%
\bm{r}).  \label{basis}
\end{equation}%
The notation $F^{\lambda }(\mathbf{r})$ alludes to the eigenvalues $\lambda $
that are related to the norm of these functions and which are centered at
a midpoint between the atoms. In LCAO one enumerates the atomic orbitals $f^{a}(\bm{r})$
with a global index $a$. Here we also enumerate the set of dominant functions of all
pairs with a single index $\lambda$. With such an enumeration, the relation
(\ref{basis}) then becomes true for arbitrary products of orbitals and with
a vertex $V_{\lambda }^{ab}$ which is sparse.
Indeed, for arbitrary orbitals $a,b$ the vertex $V_{\lambda }^{ab}$ is
non zero only for those products $F^{\lambda }(\bm{r})$ that belong to that pair of
atoms which is associated with the orbitals $a,b$.

The reduction in the number of functions occurs by ignoring the (many)
functions $F^{\lambda}(\bm{r})$ that belong to eigenvalues that are below a chosen
threshold $\lambda_{\min}$. Empirically, the norm of the omitted products vanishes
exponentially with respect to the number of basis functions retained.
Although the convergence of the physical results with respect to the cutoff
$\lambda_{\min}$ must be checked, this convergence poses no problem.

Combining eqs. (\ref{bilinear}, \ref{basis}), we may express the density operator as 
\begin{equation}
n(\bm{r},t)=\psi^{+}(\bm{r},t)\psi(\bm{r},t)=\sum_{a,b,\mu}F^{\mu}(\bm{r})
c_{a}^{+}(t)V_{\mu}^{ab}c_{b}(t).  \label{expansion_of_density}
\end{equation}%
Inserting this representation of the density into eq. (\ref{chi_product_green}),
we find a representation of the density correlator as a sum of products
of dominant functions 
\begin{equation}
\mathrm{i}\chi_{0}(\bm{r},\bm{r}^{\prime},t-t^{\prime})=\sum_{\mu,\nu
}F^{\mu}(\bm{r}) \ \chi_{\mu\nu}^{0}(t-t^{\prime})\ F^{\nu }(\bm{r}%
^{\prime}).  \label{tensor}
\end{equation}%
The entries of the matrix $\chi_{\mu\nu }^{0}$ are correlators of bilinears
of local fermions

\begin{eqnarray}
\mathrm{i}\chi_{\mu\nu}^{0}(t-t^{\prime}) &=&\sum_{iklm}\langle 0| T \{
c_{i}^{+}(t) V_{\mu}^{ik} c_{k}(t) \cdot c_{l}^{+}(t) V_{\nu}^{lm}
c_{m}(t^{\prime}) \} | 0\rangle  \label{tensor_of_correlators} \\
&=&\mathrm{Tr}(V_{\mu}G(t-t^{\prime})V_{\nu}G(t^{\prime}-t)).  \nonumber
\end{eqnarray}

\begin{figure}[htb]
\centerline{\includegraphics[width=9cm,clip]{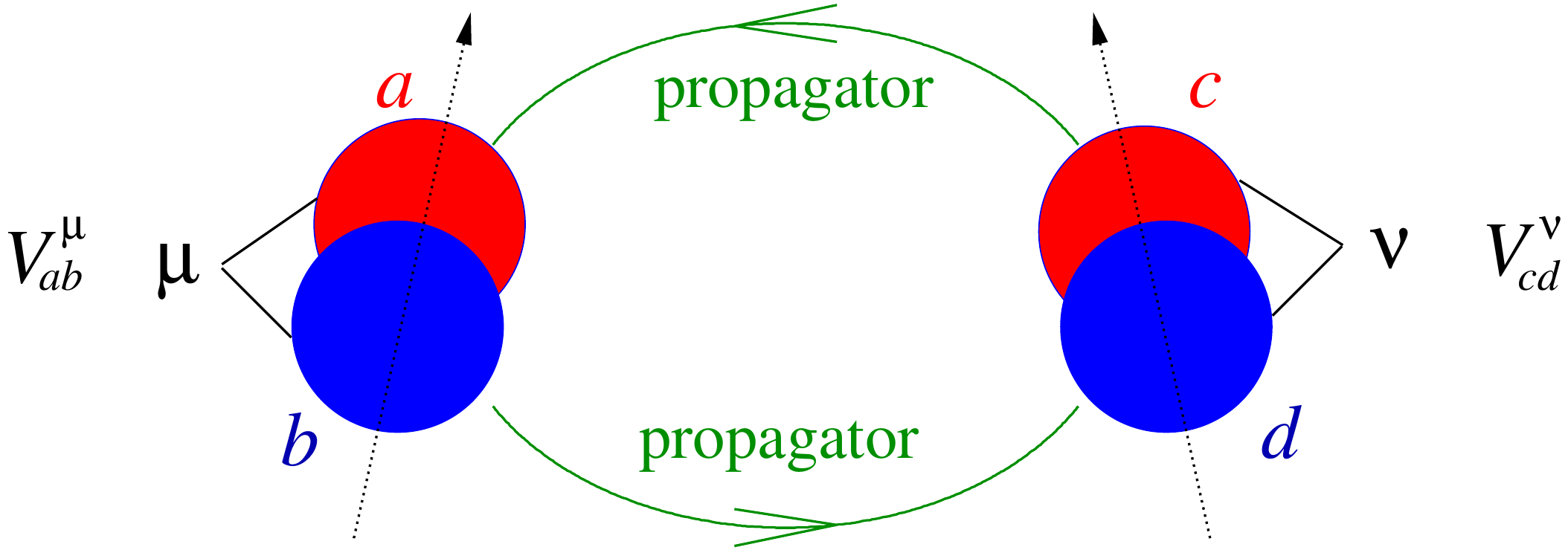}}
\caption{Particle hole graph for $\protect\chi_0$ in a basis of dominant
products. The vertex $V_{ab}^{\protect\mu}$ connects pairs of orbitals $a,b$
to a dominant product $\protect\mu$. The propagators connect the orbitals
within the orbital quadruplet. For a given pair of dominant products $(%
\protect\mu,\protect\nu)$, there are $O(N^0)$ orbitals to be summed over.
Therefore, the total computational effort scales as $O(N^2N_{\protect\omega})
$.}
\label{f:particle--hole}
\end{figure}

In this equation, the explicit expression in terms of Green's functions was
found again with the help of Wick's theorem. Equation (\ref%
{tensor_of_correlators}) describes the creation, propagation and subsequent
annihilation of a particle hole pair, see figure \ref{f:particle--hole}. The
figure shows why the construction of $\chi_{\mu\nu }^{0}$ requires $%
O(N^{2}N_{\omega })$ operations. There are $O(N)$ dominant products for the
entire molecule, and there are a total of $O(N^2)$ pairs of such products.
Due to the locality of the vertex $V_{\mu}^{ab}$, there are, for each pair,
of order $O(N^{0})$ electron propagators to be summed over. Finally, the
calculation must be done for $N_{\omega }$ frequencies.

\subsubsection*{Finding the spectral function of $\protect\chi_{\protect\mu%
\protect\nu}^{0}(\protect\omega )$}

It would be a mistake to determine $\chi_0$ directly, on the basis of
eq.~(\ref{tensor_of_correlators}), by brute force computation. At equal times,
the electron propagator has a discontinuity which hampers such an approach. 
Instead, it is better to relate the density response $\chi_0$ and the
electron propagators $G$ indirectly via their spectral functions and to
construct $\chi_{\mu\nu}^{0}$ from its spectral density at the end.

The Fourier transform of the causal (rather than the time ordered) form of
$\chi_{\mu\nu}^{0}$ is analytic in the cut complex plane. Therefore, it
should have the following Cauchy type spectral representation

\begin{equation}
\chi_{\mu\nu}^{0}(\omega+\mathrm{i}\varepsilon) =-\frac{1}{\pi }\int_{-\infty
}^{\infty } \frac{\mathrm{Im}\chi_{\mu\nu}^{0}(s)ds}{\omega +\mathrm{i}%
\varepsilon-s}.  \label{dispersion_integral}
\end{equation}%
Once we know that such a representation should exist, it is easy to identify
the spectral density by combining
eqs.~(\ref{gf},\ref{sf},\ref{tensor_of_correlators}).
After a brief calculation, we obtain the following result
\begin{eqnarray}
\chi_{\mu\nu }^{0}(\omega )&=&\int_{-\infty}^{\infty }
d\lambda\ \frac{a_{\mu\nu }(\lambda )}{\omega-\lambda+\mathrm{i}\varepsilon};
\label{starting_point} \\
a_{\mu\nu}(\lambda) &=&\iint_{0}^{\infty}d\sigma d\tau\, \sum_{abcd}
\left[V_{\mu}^{ab}\rho^{+}_{bc}(\sigma)V_{\nu}^{cd}\rho^{-}_{ad}(-\tau)\right] \delta (\sigma
+\tau-\lambda )\text{ for }\lambda>0.  \label{convolution}
\end{eqnarray}
The first line shows that the response matrix $\chi _{\mu \nu }^{0}(\omega )$
can be computed from the spectral function $a_{\mu \nu }(\lambda )$ by
taking a convolution which requires $N_{\omega }\log (N_{\omega })$
operations when done by fast Fourier methods. The second line shows that
this spectral function is a weighted convolution of particle like ($+$) and
hole like ($-$) spectral densities (\ref{sf}). As explained above after
equation (17) and in figure 1, the internal indices involved in the trace of
the second equation run only over $O(N^0)$ indices because the vertex
$V_{\mu}^{ab} $ is sparse. Computationally, it is convenient to form new spectral
functions 
$\sum_{b}V_{\mu }^{ab}\rho _{bc}^{+}(\sigma )$ and 
$\sum_{d}V_{\nu }^{cd}\rho _{ad}^{-}(\tau )$ which, thanks to the sparsity
of $V_{\mu }^{ab}$, costs $O(N^{2}N_{\omega })$ operations.

\section{Computation of $\protect\chi_{0}$ from electronic spectral densities}

\label{s:3}

To compute the convolutions in eq. (\ref{convolution}) efficiently, we make
extensive use of the fast Fourier transform that does such convolutions in $%
O(N_{\omega }\log N_{\omega })$ operations for $N_{\omega }$ frequency
points \cite{NumericalRecipes}. In this section, we will explain (i)~how to
discretise the electronic spectral density on a frequency lattice and
(ii)~how to evaluate the spectral integral over the infinite frequency
interval in eq.~(\ref{starting_point}).

\subsubsection*{Discretizing the spectral density}

We will discretize the electronic spectral densities in eq. (\ref{sf}) in a
window $(-\omega _{\max },\omega _{\max })$ and on a grid with spacing $%
\Delta \omega =\frac{\omega _{\max }}{N_{\omega}}$. To hide the effect
this might have on $\chi _{\mu \nu }^{0}(\omega )$ we will later broaden the
spectral resolution by adding a small imaginary part $\mathrm{i}\varepsilon$
to the frequency. Let the grid points be defined as follows
\begin{equation}
\omega _{n}=\frac{n}{N_{\omega}}\omega _{\max },\Delta \omega =\frac{\omega
_{\max }}{N_{\omega}},n=1-N_{\omega}\ldots N_{\omega}-1.  \label{fine-grid}
\end{equation}

\begin{figure}[tbh]
\centerline{%
\includegraphics[width=7cm,clip]{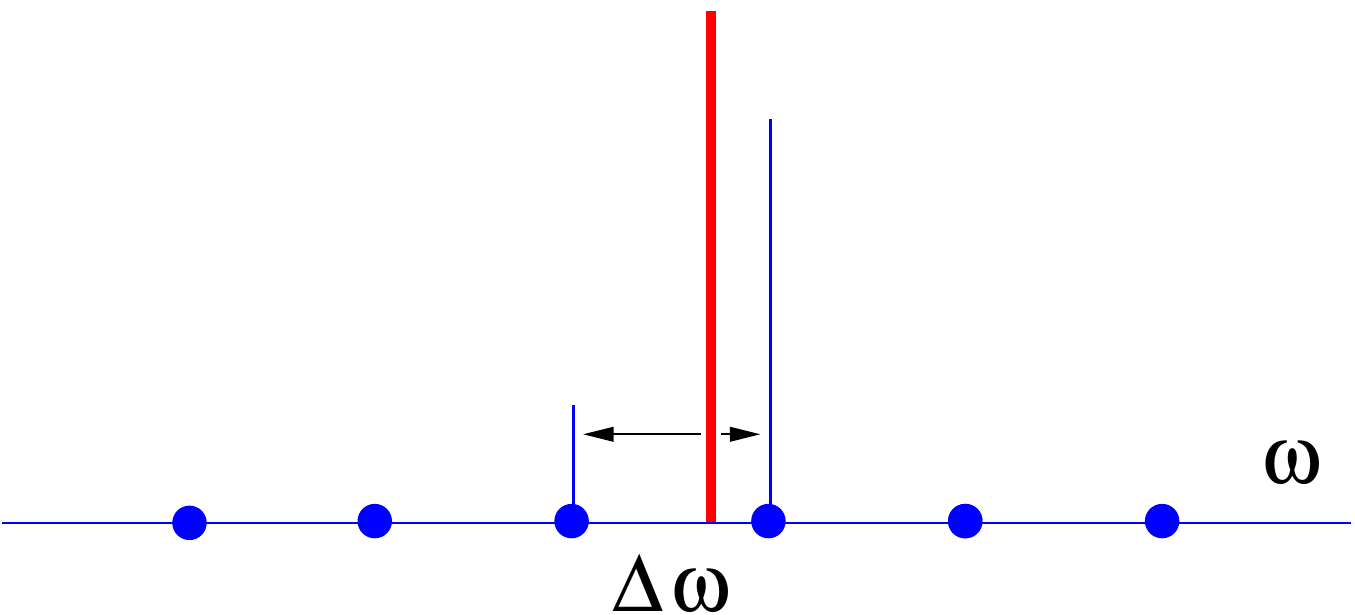}}
\caption{Redistribution of spectral weight on the frequency mesh.}
\end{figure}
Consider an eigenenergy $E$ that belongs to the frequency window 
$-\omega_{\max }$ $<E<\omega _{\max }$ 
\footnote{Energies are measured with respect to a ``Fermi energy'' \
-- halfway between the LUMO and HOMO states} and
which is located between two successive mesh points $(\omega _{n},\omega
_{n+1})$. We distribute the spectral weight $X_{a}^{E}X_{b}^{E}$ to
the neighboring frequencies $\omega _{n},\omega _{n+1}$ in a way that
conserves (i)~the total spectral weight and (ii)~it's center of mass by
using the following weight factors $p_{n},p_{n+1}$ 
\begin{equation}
p_{n}=\frac{\omega _{n+1}-E}{\Delta \omega },\ \ p_{n+1}=1-p_n.  \label{weights}
\end{equation}%
Alternatively, one may also minimize the norm of the difference $\Delta $
between the pole at $\omega=E$ and its representation by poles at the two neighboring
frequencies on the lattice 
\begin{eqnarray}
\Delta (\omega )&=&\frac{p_{n}}{\omega -(\omega _{n}+\mathrm{i}\varepsilon )}+%
\frac{p_{n+1}}{\omega -(\omega _{n+1}+\mathrm{i}\varepsilon )}-\frac{1}{%
\omega -(E+\mathrm{i}\varepsilon )}\equiv \sum_{i=n,n+1,0}\frac{p_{i}}{%
\omega -\omega_{i}+\mathrm{i}\varepsilon},\\
p_{0}&=&-1,\  p_{n+1}=1-p_n, \ \omega_0 = E. \nonumber \label{error_norm}
\end{eqnarray}%
There is a simple expression for the norm of this error that can be obtained
by contour integration
\begin{equation}
||\Delta ||^{2}=\frac{1}{2\pi }
\int_{-\infty }^{\infty }|\Delta(\omega)|^{2}d\omega =
2\varepsilon
\sum_{i,k=n,n+1,0}
\frac{p_{i} p_{k}}{(\omega_{i}-\omega_{k})^2+4\varepsilon^2}.
\end{equation}%
With $\varepsilon \gtrsim \Delta \omega $, the coefficient $p_{n}$,
that minimizes the error norm, varies almost linearly between
$0$ and $1$ as a function of $\frac{E}{\Delta \omega }$
and differs little from eq.~(\ref{weights}).
As the errors are of the same order in both cases, we use the first and simpler method 
according to eq.~(\ref{weights}). This part of the calculation 
actually requires $O(N^{3})$ operations, but the prefactor is very
small -- the discretization of the spectral data of benzene takes about a
second on a current personal computer.

\begin{figure}[htb]
\centerline{
\includegraphics[width=9cm,viewport=50 40 400 290,clip]{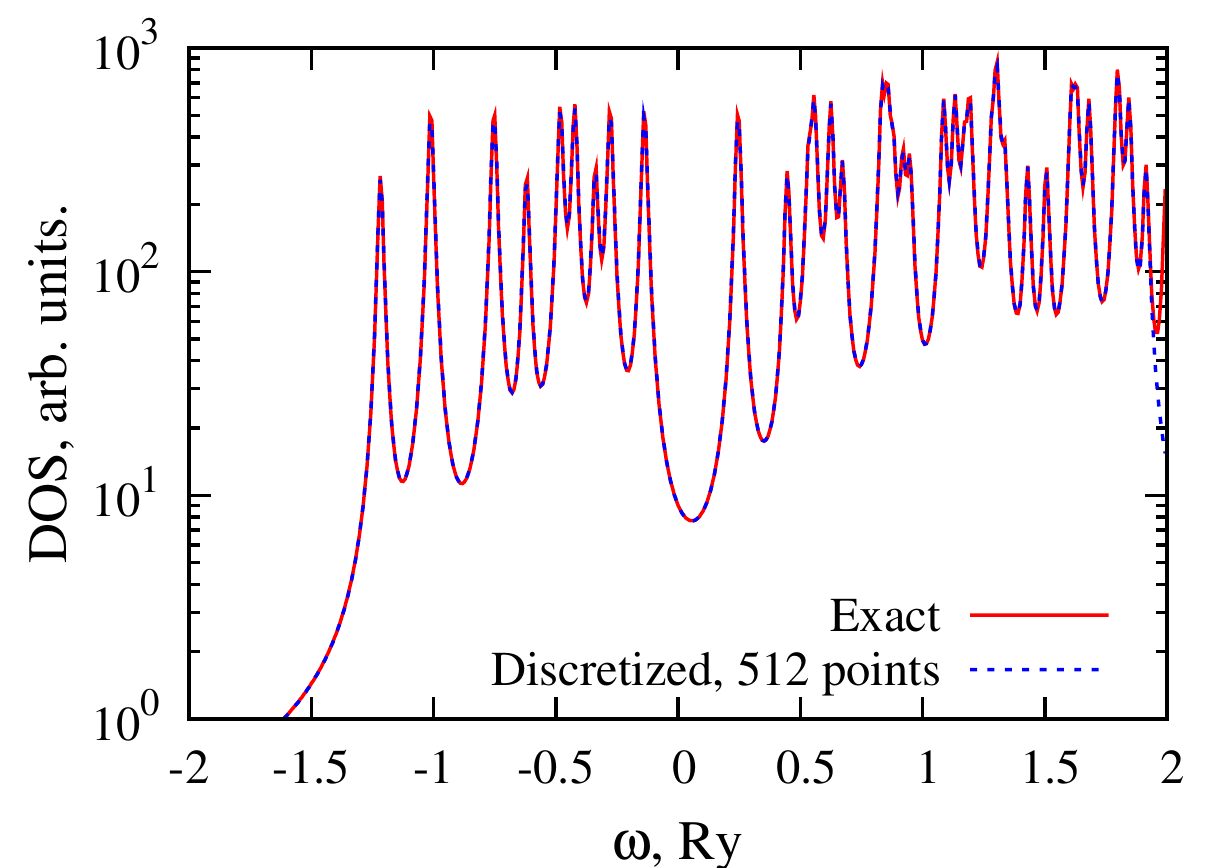}
\includegraphics[width=9cm,viewport=50 40 400 290,clip]{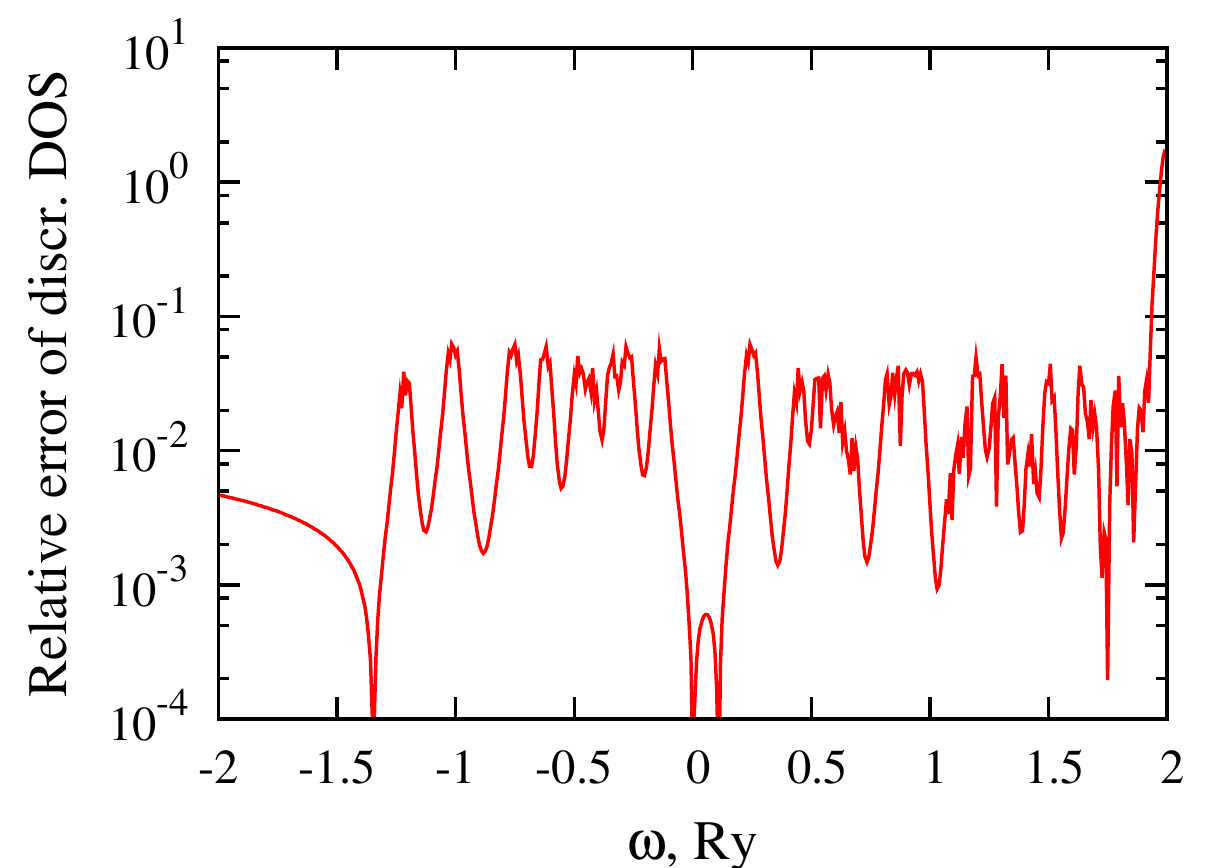}}
\caption{The (noninteracting) density of states (DOS) of benzene. The exact
DOS $\sum_E(\protect\omega - E+\mathrm{i}\protect\varepsilon)^{-1}$ is
computed with eigenvalues $E$ obtained using the Siesta package \protect\cite%
{siesta}. Default settings were used in the Siesta run. The discretized DOS
is computed in the frequency window 
$-\protect\omega_{\max}< \protect\omega<\protect\omega_{\max}$, $\protect\omega_{\max}=2$
Rydberg with 
$N_{\protect\omega}=512$ data points. $\protect\varepsilon$ is chosen to be 
$1.5 \Delta\protect\omega$, where the discretisation spacing is 
$\Delta\protect\omega=2\protect\omega_{\max}/N_{\protect\omega}$. }
\label{f:dos}
\end{figure}

To judge the quality of this discretisation, we compute the density of
states $-\frac{1}{\pi}\mathrm{Tr} (S\, \mathrm{Im}\,G(\omega+\mathrm{i}%
\varepsilon ))$ in the case of benzene, within a window of frequencies
(i)~by direct calculation from the exact Green's function and (ii)~after
redistributing the spectral weights. Figure \ref{f:dos} shows that the two
densities of states differ very little. The good agreement between the two
densities vindicates our discretization procedure.

\subsubsection*{The need for a second spectral window}

According to eq.~(\ref{starting_point}) we must find an integral over the
full spectral range $(0,\Omega_{\max} )$ even if we want spectroscopic results only
for low frequencies $\omega \leq \omega _{\max }$. We resolve this
difficulty by decomposing the integral in eq.~(\ref{starting_point}) as
follows 
\begin{eqnarray}
\chi_{\mu \nu }^{0}(\omega ) &=&\left( \int_{0}^{\omega_{\max}}+
\int_{\omega_{\max }}^{\Omega_{\max}}\right) a_{\mu \nu }(\lambda )\left( 
\frac{1}{\omega +\mathrm{i}\varepsilon -\lambda }-\frac{1}{\omega +\mathrm{i}%
\varepsilon +\lambda }\right) d\lambda   \label{spectralrepresentation} \\
&\equiv &\chi _{\mu \nu }^{0\text{, resonant}}(\omega )+\chi _{\mu \nu }^{0%
\text{, nonresonant}}(\omega ).  \nonumber
\end{eqnarray}%
The first term $\chi _{\mu \nu }^{0\text{, resonant}}$ \ in this
decomposition has resonant structure because $\omega $ and $\lambda $ may
coincide in the denominator of its integrand. By contrast, the integrand in
the expression for $\chi_{\mu \nu }^{0\text{, nonresonant}}$ is regular and
therefore this function has much less structure. In the resonant part, we
must allow for sufficiently many grid points to capture the features of the
spectral density. In the nonresonant part, we determine the spectral
density for the full range of Kohn--Sham eigenvalues, and for simplicity, we
use the same number of gridpoints $N_{\omega}$. However, we need the
resulting response function 
$\chi_{\mu \nu }^{0,\text{ nonresonant}}(\omega)$ only in the frequency interval
$(0,\omega_{\max })$ where we find its
values on the corresponding grid points (\ref{fine-grid}) by interpolation. 

\begin{figure}[htb]
\includegraphics[width=8cm,viewport=50 40 400 290,clip]{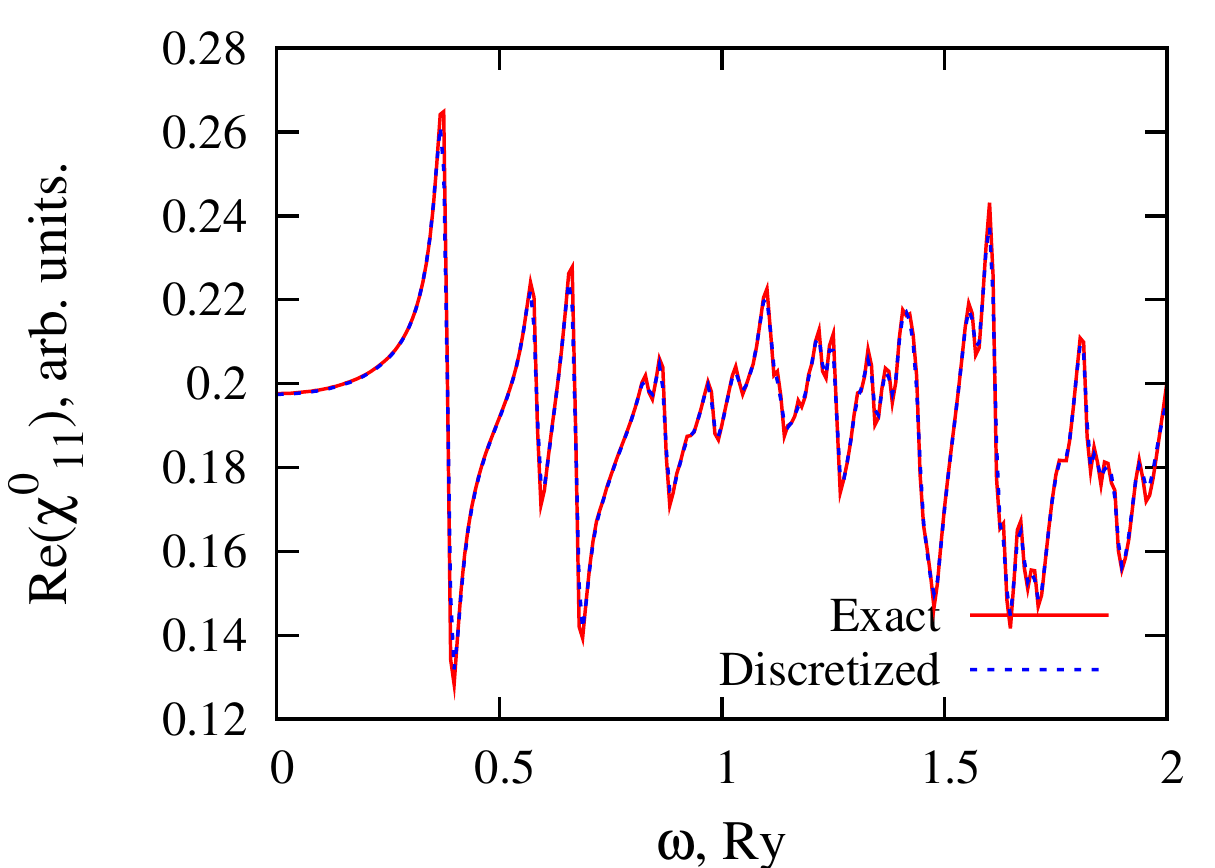} %
\includegraphics[width=8cm,viewport=50 40 400 290,clip]{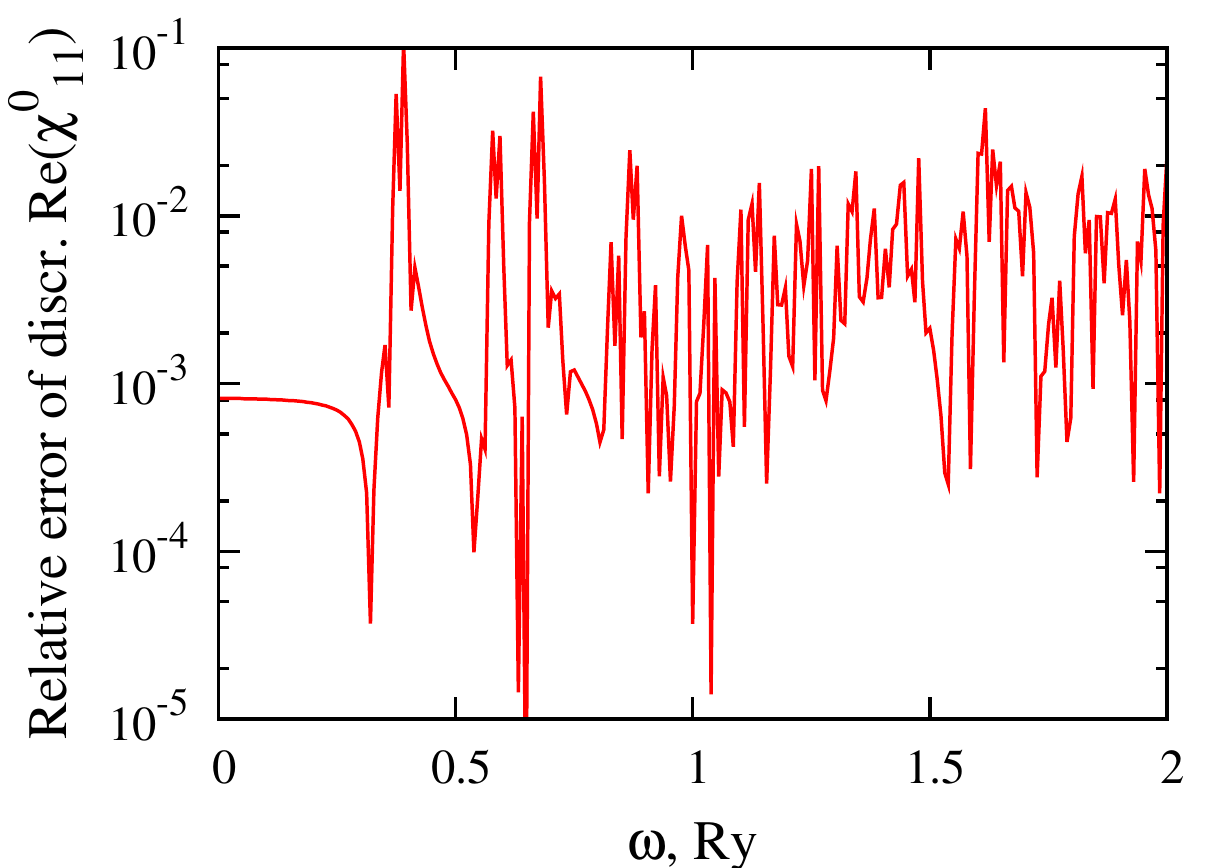}
\caption{An element of the response function $\protect\chi^0_{\protect\mu%
\protect\nu}$ of benzene. The exact response is computed according to eq. (%
\protect\ref{exact_response_matrix}). The Kohn--Sham eigenstates were
generated using the Siesta package \protect\cite{siesta} with default
settings. The discretized response function is computed in the frequency
window $\protect\omega<\protect\omega_{\max}$, $\protect\omega_{\max}=2$
Rydberg with $N_{\protect\omega}=512$ data points. $\protect\varepsilon$ is
chosen to be $1.5 \Delta\protect\omega$, where the discretisation spacing is 
$\Delta\protect\omega=2\protect\omega_{\max}/N_{\protect\omega}$.}
\label{f:chi0_11}
\end{figure}

To judge the quality of $\chi _{\mu \nu }^{0}(\omega )$ constructed in this
way, we make use of the exact expression (\ref{conventional}) for the
response function. The corresponding response matrix $\chi _{\mu \nu }^{0%
\text{ exact}}(\omega )$ can be obtained by expressing $\varphi _{E}(\bm{r}%
)\varphi _{F}(\bm{r})$ in eq. (\ref{conventional}) in terms of dominant
functions. Using eqs. (\ref{conventional},\ref{lcao},\ref{basis}) we obtain
\begin{equation}
\chi _{\mu\nu }^{0,\mathrm{exact}}(\omega )=\sum_{E,F,E\cdot F<0;abcd}\frac{%
n_{F}-n_{E}}{\omega -(E-F)-\mathrm{i}\varepsilon (n_{E}-n_{F})}
\left(X_{a}^{E}V_{\mu}^{ab}X_{b}^{F}\right) \left( X_{c}^{F}V_{\nu
}^{cd}X_{d}^{E}\right).  \label{exact_response_matrix}
\end{equation}%
Actually, the exact response matrix $\chi_{\mu\nu }^{0,\mathrm{exact}%
}(\omega )$ requires $O(N^{4}N_{\omega })$ operations and it takes too long
to compute for other than very small molecules. Nonetheless, the exact expression
(\ref{exact_response_matrix}) is well suited as a test provided we use it only
for fixed entries $\mu,\nu$. Figure~\ref{f:chi0_11} indicates that the error
is well controlled and vindicates our ``two windows technique'' for
constructing $\chi_{\mu\nu}^{0}(\omega)$.

\begin{figure}[tbp]
\centerline{
\includegraphics[width=8.0cm,viewport=50 40 400 290,clip]{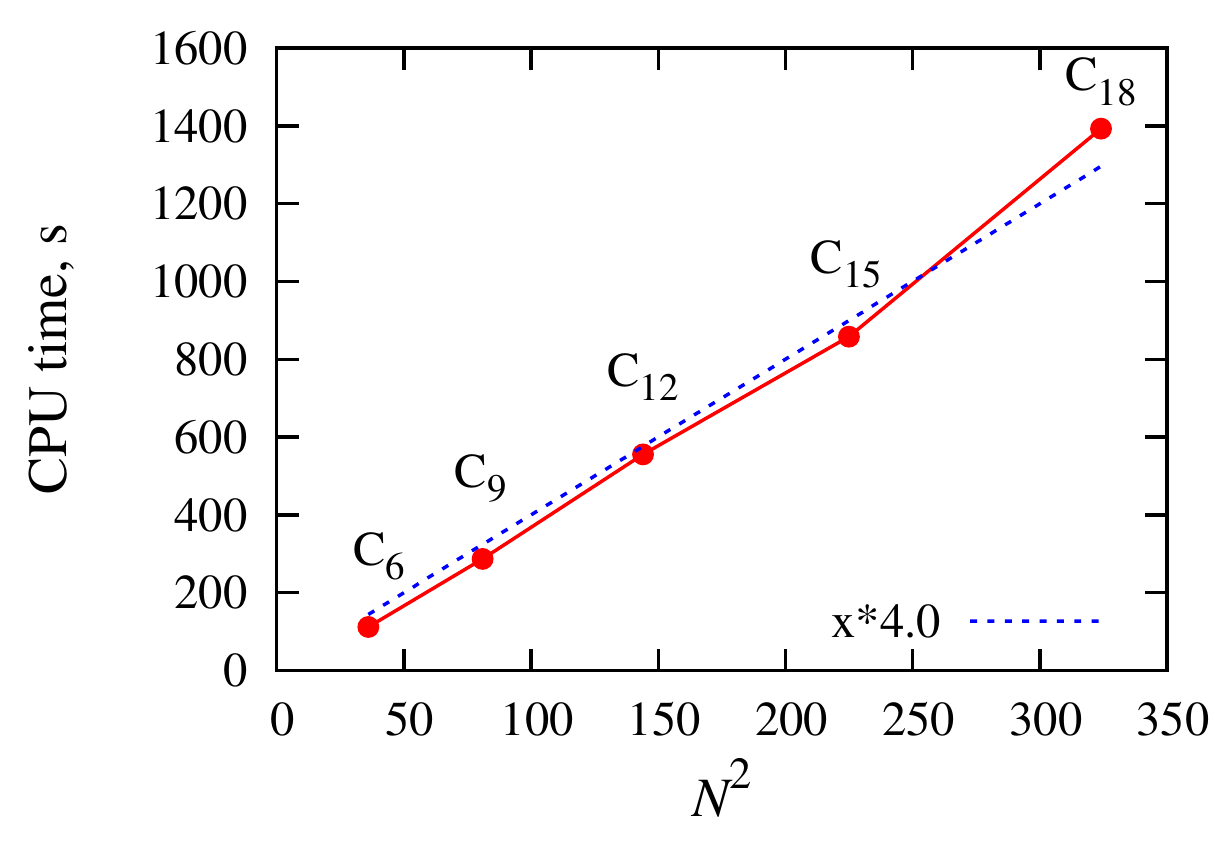}}
\caption{CPU time for computing $\protect\chi_0$ as a function of the number
of atoms $N$ in the case of carbon chains.}
\label{f:scaling-sqr-natoms-time}
\end{figure}

We argued before that the total computational cost of our method scales as $%
O(N^{2}N_{\omega })$ and we believe that this scaling is the best that can
be achieved for the noninteracting response function $\chi _{0}$. In order
to confirm this scaling, we computed the noninteracting response $\chi _{0}$
for a number of carbon chains, measured the wall clock time and represented
it in figure \ref{f:scaling-sqr-natoms-time}. The scaling law is slightly
disturbed, probably due to the high memory requirement of our algorithm in
the case of the $C_{18}$ chain.

\section{Testing $\protect\chi_{0}$ in the calculation of molecular spectra}

\label{s:4}

\subsubsection*{The Petersilka--Gossmann--Gross equation}

In the previous section, we gave a first test of our construction of $%
\chi^{0}_{\mu\nu}$ by comparing with an exact result. Here we will further
test $\chi^{0}_{\mu\nu}$ by using it to compute molecular spectra from the
Petersilka--Gossmann--Gross equations of TDDFT linear response \cite{GrossPetersilka} 
\begin{eqnarray}
\chi^{-1}(\bm{r},\bm{r}^{\prime},\omega) &=&\chi_{0}^{-1}(\bm{r},\bm{r}%
^{\prime},\omega)- f_{\mathrm{H}}(\bm{r},\bm{r}^{\prime})-f_{%
\mathrm{xc}}(\bm{r},\bm{r}^{\prime}) ;  \label{GrossPolarisation} \\
P_{ik}(\omega ) &=&\int d\bm{r}\,d\bm{r}^{\prime}\ \bm{r}_{i}\chi(\bm{r},%
\bm{r}^{\prime},\omega)\bm{r}_k^{\prime}.
\label{dipole-polarizability-tensor}
\end{eqnarray}
The results will be compared with spectra obtained using Casida's equations 
\cite{Casida}. The Petersilka--Gossmann--Gross equations are a consequence
of a generalisation of the Kohn--Sham equations of the electron
gas~\cite{KohnShamEquation} to time dependent electron densities

\begin{equation}
V_{\mathrm{KS}}(\bm{r},t)=V_{\mathrm{ext}}(\bm{r},t)+
V_{\mathrm{H}}(\bm{r},t)+V_{\mathrm{xc}}(\bm{r},t).  \label{time_dependent_Kohn_Sham}
\end{equation}%
Here $V_{\mathrm{KS}}(\bm{r},t)$ is the potential that assures a prescribed
density $n(\bm{r},t)$ of the noninteracting Kohn--Sham reference electrons, $%
V_{\mathrm{H}}(\bm{r},t)=2\int \frac{n(\bm{r}^{\prime },t)}{|\bm{r}-\bm{r}%
^{\prime }|}d\bm{r}^{\prime }$ (the factor 2 is from spin) and 
$V_{\mathrm{xc}}(\bm{r},t)$ is the exchange correlation potential. All
quantities in this equation depend on the electronic
density $n(\bm{r},t)$. We differentiate both sides with respect to
this density and, upon using 
$\chi _{0}=\frac{\delta n}{\delta V_{\mathrm{KS}}}$, 
$\chi =\frac{\delta n}{\delta V_{\mathrm{ext}}}$, we obtain
equation~(\ref{GrossPolarisation}) with the following kernels 
\begin{equation}
f _{\mathrm{H}}=2\frac{\delta (t-t^{\prime })}{|\bm{r}-\bm{r}^{\prime }|%
}\text{ \ and \ \ }f _{\mathrm{xc}}=\frac{\delta V_{\mathrm{xc}}(\bm{r}%
,t)}{\delta n(\bm{r}^{\prime },t^{\prime })}, \ f_{\mathrm{Hxc}} = f_{\mathrm{H}}+f_{\mathrm{xc}}.
\label{kernels}
\end{equation}%
We make the conventional ``adiabatic'' assumption that
$f _{\mathrm{xc}}$ has no memory and that it depends only on the instantaneous
electron density. Therefore, both $f _{\mathrm{H}}$ and
$f_{\mathrm{xc}}$ are local in time and their Fourier transforms
are frequency independent.

To write the Petersilka-Gossmann-Gross equations in our basis of dominant
functions, we start with the integral form of this equation \cite{GrossPetersilka}:
\begin{equation*}
\chi(\bm{r},\bm{r}',\omega) =\chi_{0}(\bm{r}, \bm{r}')+
\int d\bm{r}''d\bm{r}'''\ 
\chi_{0}(\bm{r}, \bm{r}'', \omega) f_{\mathrm{Hxc}}(\bm{r}'', \bm{r}''') \chi(\bm{r}''', \bm{r}', \omega).
\end{equation*}
In our basis of products and with the parametrizations 
$\chi(\bm{r},\bm{r}',\omega )=\sum_{\mu \nu }F^{\mu }(\bm{r})
\chi_{\mu\nu }(\omega )F^{\nu}(\bm{r}')$ 
and $\chi_{0}(\bm{r},\bm{r}',\omega)=
\sum_{\mu \nu}F^{\mu }(\bm{r})\chi_{\mu\nu }^{0}(\omega)F^{\nu }(\bm{r}')$,
this Dyson equation takes the following form%
\begin{eqnarray}
\chi _{\mu \nu }(\omega ) &=&\chi _{\mu \nu }^{0}(\omega )+
\sum_{\alpha\beta}
\chi _{\mu \alpha}^{0}(\omega)f_{\mathrm{Hxc}}^{\alpha\beta}\chi _{\beta \nu }(\omega ).
\label{Matrix_Dyson}
\end{eqnarray}%
In the last section, we computed $\chi_{0}=\frac{\delta n}{\delta V_{%
\mathrm{KS}}}$. In the next subsections we will compute the kernels $f_{%
\mathrm{H}}$, $f_{\mathrm{xc}}$ and the polarizability $P_{ik}(\omega )$.

\subsubsection*{Computing the kernels $f_{\mathrm{H}}$ and $%
f_{\mathrm{xc}}$ in a basis of dominant products}

In the basis of dominant products, the Hartree part of the kernel reads 
\begin{equation}
f _{\mathrm{H}}^{\mu \nu }=\int d\mathbf{r}d\mathbf{r}^{\prime }\
F^{\mu }(\mathbf{r})\frac{1}{|\mathbf{r}-\mathbf{r}^{\prime }|}F^{\nu }(%
\mathbf{r}^{\prime }).
\end{equation}%
For the present discussion to be reasonably self contained, we must give
more details on the structure of the dominant products \cite{DF}. 
As seen previously in section~\ref{s:2}, the dominant products
were constructed in the context of the LCAO method where molecular
orbitals are expanded as in eq.~(\ref{lcao}).
Therefore, orbital products and the dominant products constructed
from them have either spherical or only axial symmetry depending on
whether the two atoms that give rise to them coincide or not.
Technically, the products are represented as expansions in spherical
harmonics (in appropriate local coordinates, in the case of bilocal products)
about a midpoint between the two atoms that form the pair.

The Hartree kernel $f_{\mathrm{H}}^{\mu\nu }$ involves two
products $F^{\mu }(\bm{r})$, $F^{\nu }(\bm{r}')$ that belong, generally, to
two distinct pairs of atoms with their own axial or spherical symmetry and
local coordinates. With the help of Wigner's rotation matrices $%
d_{mm^{\prime}}^{j}$ \cite{Blanco-etal:1997} the two distinct products can
be referred to a single reference frame. In the end, the Hartree kernel
is reduced to a sum of conventional two center integrals 
\begin{equation}
\int d\bm{r}_{1} d\bm{r}_{2}\ g_{j_{1}m_{1}}(\bm{r}_{1}-\bm{c}_{1}) \frac{1}{%
|\bm{r}_{1}-\bm{r}_{2}|} g_{j_{2}m_{2}}(\bm{r}_{2}-\bm{c}_{2}),
\end{equation}%
where the elementary functions $g_{jm}(\bm{r})=g_j(r) S_{jm}(\bm{r})$ are
explicitly of spherical symmetry \footnote{%
We use real spherical harmonics $S_{jm}(\bm{r})$ in our calculation to
improve the performance.}.

The calculation of such conventional two center integrals is conveniently
done in momentum space and using Talman's fast Bessel transform
\cite{FastBessel} to relate real space orbitals to their Fourier images.

Due to the finite support of the dominant products, the Hartree kernel
must be integrated explicitly only for a subset of $O(N)$ pairs of mutually
overlappinging dominant products. The Coulomb interaction of the remaining
nonoverlapping pairs of products can be calculated exactly and cheaply as an
interaction between their multipoles.

By contrast, the remaining kernel $f_{\mathrm{xc}}$ is a
3-dimensional integral in the local density approximation%
\begin{equation}
f _{\mathrm{xc}}^{\mu \nu }=\int d\mathbf{r}\ F^{\mu }(\mathbf{r})\frac{%
dV_{\mathrm{xc}}}{dn}F^{\nu }(\mathbf{r}).  \label{sigma-xc}
\end{equation}%
There are only $O(N)$ such matrix elements to calculate because the basis
functions $F^{\mu }(\mathbf{r})$ have finite support. In the general case,
the integration domain is an overlap of two distinct lenses because the
support of each dominant product is an overlap of two spheres. In spite of
this, we used a simple numerical integration in spherical coordinates as an
easy alternative to more elaborate integration techniques, with the center
of spherical coordinates on the midpoint between the two centers, each
associated with a dominant product. The integration over solid angle is done
via Lebedev's method \cite{Lebedev-theory-and-program} and integration over
the radial coordinates is done by the Gauss--Legendre method. By default, we
use 86 grid points in Lebedev integration and 24 grid points in
Gauss--Legendre integration.

\begin{figure}[htb]
\centerline{\includegraphics[width=6cm,height=7cm,viewport=325 135 480
325,angle=270,clip]{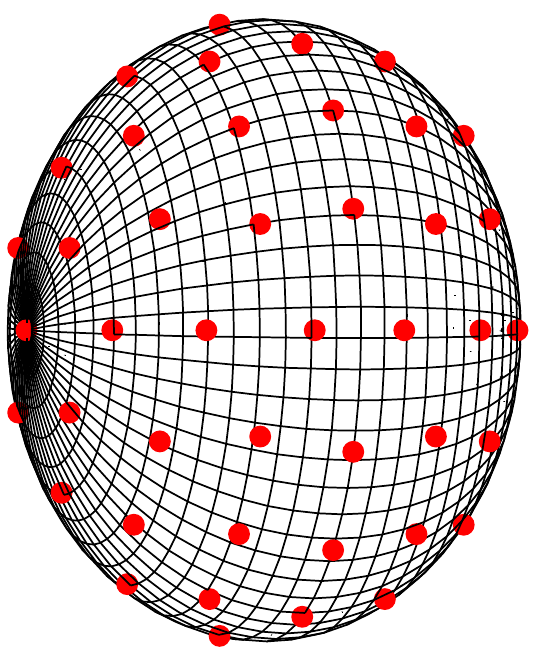}}
\caption{The nodes of a Lebedev grid with 86 points. With this grid we can
exactly integrate any linear combination of spherical harmonics up to
angular momentum $j=15$. }
\label{f:lebedev-grid}
\end{figure}

\subsubsection*{Finding the molecular polarizability}

Using the matrix form of the density response $\chi$ from eq.~(\ref{Matrix_Dyson}),
we find the interacting polarizability $P_{ik}(\omega )$ %

\begin{eqnarray}
P_{ik}(\omega ) &=& \bm{d}_{i} \chi \bm{d}_{k} =\bm{d}_{i} \frac{1}{%
1-\chi_{0}(\omega )f_{\mathrm{Hxc}} } \chi_{0}(\omega)\bm{d}_{k} ,
\label{polarizability} \\
\bm{d}_i^{\mu}&=&\int d\bm{r} \ \bm{r}_i\, F^{\mu }(\bm{r}).  \nonumber
\end{eqnarray}%
Here $\bm{d}^{\mu}$ is a vector of dipole moments that is associated with
the dominant products $F^{\mu }(\bm{r})$. We may compute the polarizability
(\ref{polarizability}) by matrix inversion or,
alternatively, by solving $N$ linear equations in $N$ variables
to find $\chi\, \bm{d}_k$. 
Either method requires $O(N^{3}N_{\omega })$ operations
which is worse than the $O(N^{2}N_{\omega })$ scaling in the computation of $%
\chi _{0}$. On the other hand, equation (\ref{polarizability}) shows that
the polarizability does not see the full matrix $\chi $ but only its (low
rank) projection onto the dipole moments $\bm{d}$. Fortunately, iterative
Lanczos--Krylov methods \cite{Saad} are capable of finding such projections of the
inverse of a matrix in $O(N^{2})$ operations.

To find the inverse of a matrix $A=1-\chi _{0}(\omega )f_{\mathrm{Hxc}}$ contracted with
two vectors $\langle L|=\bm{d}_{i}$ and $|R\rangle =\chi _{0}(\omega )\bm{d}_{i}$,
we use a biorthogonal Lanczos construction based on the two sets of Krylov spaces
$\{A^{n}|R\rangle \},\ \{\langle L|A^{n}\}$. This construction provides us
with (i) a set of orthonormal states $\langle n|m\rangle =\delta _{mn}$,
with (ii) a tridiagonal representation of $A$ and (iii) with an easily
calculable inverse of $A$ within the Krylov spaces $\{A^{n}|R\rangle \},\
\{\langle L|A^{n}\}$

\begin{equation}
A\sim \sum_{m,n}|m\rangle t_{mn}\langle n|\text{ and \ \ }A^{-1}\sim
\sum_{m,n}|m\rangle t_{mn}^{-1}\langle n|  \label{tridiagonal}
\end{equation}%
(we wrote \textquotedblleft $\sim $\textquotedblright\ because the
construction is at most asymptotic). We find the following
representation of the trace of the polarizability (relevant when
averaging over directions) 
\begin{eqnarray}
\frac{1}{1-\chi _{0}(\omega )f_{\mathrm{Hxc}} } &=&\sum |m\rangle t_{mn}^{-1}\langle
n|,  \label{Scalar_Lanczos} \\
P_{ii}(\omega ) &=&\langle L|1\rangle t_{11}^{-1}\langle 1|R\rangle
=t_{11}^{-1}P_{ii}^{0}(\omega ).  \nonumber
\end{eqnarray}%
The relation $\langle L|1\rangle \langle 1|R\rangle =\langle L|R\rangle
=P_{ii}^{0}(\omega )$ is a simple normalization condition that follows also
from the biorthogonality of the Lanczos vectors. Equation
(\ref{Scalar_Lanczos}) shows that the interaction causes the Kohn--Sham
polarizability to be multiplied by a factor that is the $(1,1)$ component of
the inverse of the matrix $t^{-1}$. With a small Krylov dimension of $%
O(N^{0})$, the calculational effort scales as $O(N^{2})$ \footnote{%
We must apply $\chi _{0}(\omega )$ and $f $ consecutively on vectors,
rather than forming the matrix $1-\chi _{0}(\omega )f $ which would
require $O(N^{3})$ operations. We assumed the dimension of the Krylov space
to be of order $O(N^{0})$, but we have not checked this in detail.}.

If the full polarization tensor $P_{ik}$ is wanted, then it is better to use
a block Lanczos procedure \cite{Saad}. We then consider the following Krylov
spaces and biorthogonalize them 
\begin{equation}
\langle L,i|_{i=1..3}A^{n}\text{ \ and \ \ }A^{n}|R,i\rangle _{i=1..3},
\end{equation}%
where $|R,i\rangle $ and $\langle L,i|$ \ represent, respectively, $%
\chi _{\mu \nu }^{0}(\omega )d_{i}^{\nu }$ and $d_{i}^{\mu }$. The
scalar representation (\ref{Scalar_Lanczos}) is now replaced by the
following block representations of $(1-\chi _{0}(\omega )f_{\mathrm{Hxc}} )^{-1}$

\begin{equation}
\frac{1}{1-\chi _{0}f_{\mathrm{Hxc}} }= \sum |m,i\rangle t_{mi,nk}^{-1}\langle n,k|.
\label{block_inverse}
\end{equation}%
Applying this to $P_{ik}(\omega )$ \ we found 
\begin{equation}
P_{ik}(\omega ) = \sum_{ab}\langle L,i|1,a\rangle\left(
t^{-1}\right)_{1a,1b}\langle 1,b|R,k\rangle.
\end{equation}%
We chose to keep the left vectors at the lowest Krylov level unchanged and
obtain $\langle 1,a|1,b\rangle=\delta_{ab}$ as a normalization condition. We
therefore find the following simple matrix relation between $P(\omega )$ and 
$P_{0}(\omega )$ 
\[
P(\omega )=\left( t^{-1}\right)_{11}P_{0}(\omega ).
\]%
The details of the block Lanczos algorithm \cite{Saad} are not given here. They are
standard and may be obtained from the authors upon request.
 
 \subsection*{Electronic excitation spectra of molecules}
 
 In the previous section, we described a numerical procedure for calculating
 the dynamical polarizability $P_{ik}(\omega )$ in $O(N^{2}N_{\omega }$)
 operations. Our implementation of this algorithm contains a number of
 computational parameters that have to be adjusted properly. For instance,
 the precision of our Lanczos method depends on the dimension of its Krylov
 space. In the examples below, a very small Krylov dimension $\lesssim 10$
 gave a polarization $P(\omega)$ with a relative error of $\lesssim 10^{-2}$.
 Other computational parameters were carefully cross checked and the results
 of some of these calculations are given in the figures \ref{f:dos} and \ref%
 {f:chi0_11}.
 
 \begin{figure}[htb]
 \centerline{
 \includegraphics[width=9cm,viewport=50 40 400 290,clip]{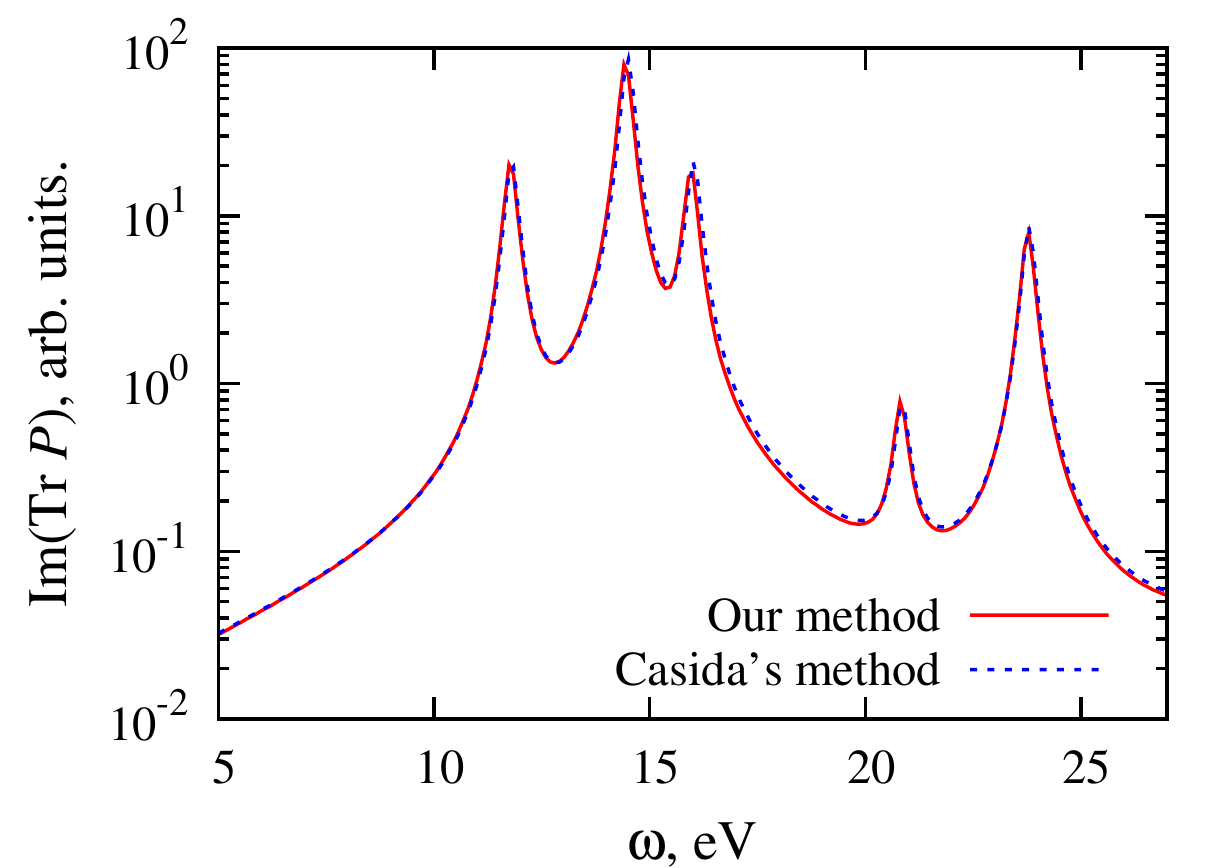}
 \includegraphics[width=9cm,viewport=50 40 400 290,clip]{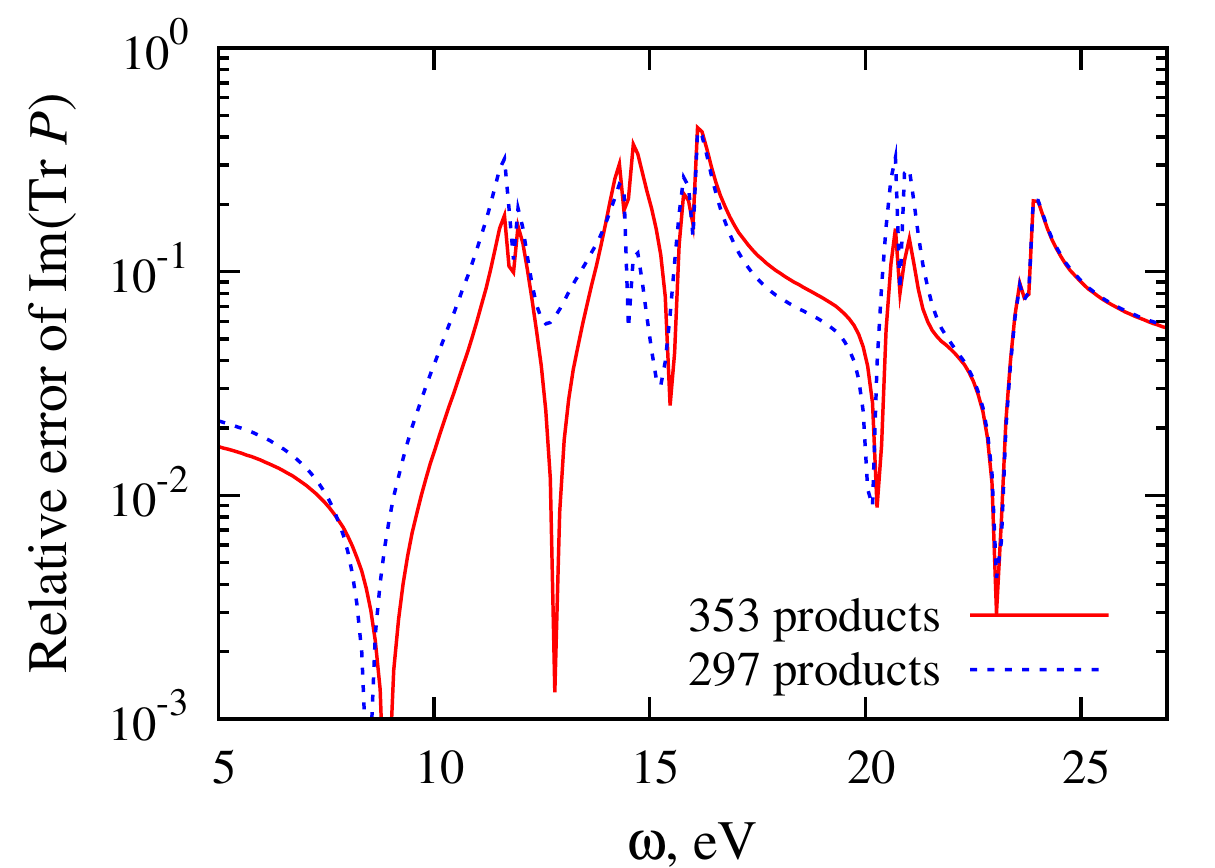}}
 \caption{The dynamical polarizability of methane calculated by our method and compared with that of
 Casida. We use deMon2k's default basis set (DZVP) and the Perdew--Zunger exchange--correlation potential.}
 \label{f:spectrum-methane-demon}
 \end{figure}
 
 In order to test the method as a whole, we compared our polarizabilities
 with those computed from Casida's equations \cite{Casida} with the help of
 the deMon2k package \cite{www-demon}. Casida's equations allow the
 determination of excitation energies $\omega_{I}$ and corresponding
 oscillator strengths $f_{I}$ and provide a dynamical polarizability that is
 parametrized as 
$$
\frac{1}{3}\mathrm{Tr}P(\omega )=\,\frac{1}{3}\sum_{I}\ \frac{f_{I}}{(\omega
 +\mathrm{i}\varepsilon )^{2}+\omega_{I}^{2}}. 
$$
We successfully compared results for several small molecules: hydrogen,
methane, methane dimer, benzene and diborane. The results of the two methods
for methane are presented in figure \ref{f:spectrum-methane-demon} where we
see a reasonable agreement. To achieve this agreement we had to discretize
the basis orbitals (contracted Gaussians in deMon2k) on our numerical grid
and import them into our code. In all cases, our results converge to
those of Casida when we enlarge our basis of dominant products. However, a
large number of dominant products is needed in order to achieve convergence.
For instance, we had to take about 360 dominant products for methane
and more than 1800 for benzene. This is due to the comparatively
large support of the Gaussian basis in deMon2k. Therefore, in the next
example, we used a basis of numerical atomic orbitals which is far more
natural for our method.

\begin{figure}[htb]
\centerline{\includegraphics[width=9cm,viewport=50 40 400 290,clip]{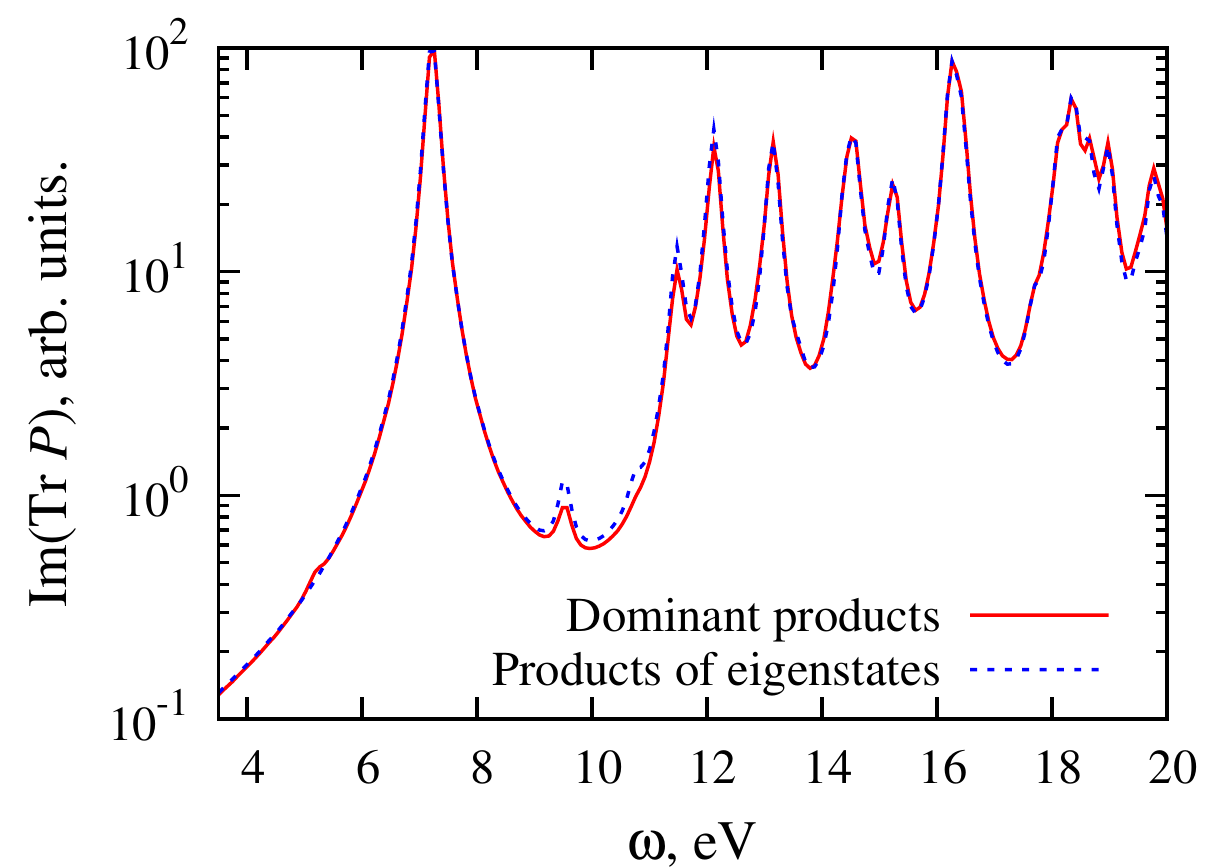}}
\caption{Spectra of benzene computed in a basis of dominant products and with the 
original products of eigenstates. Kohn--Sham eigenstates have been imported
from the Siesta package \cite{siesta}. Default settings were used in
the Siesta run: a double zeta polarized basis set (DZP) and the
Perdew--Zunger exchange--correlation potential.}
\label{f:spectrum-benzene-siesta}
\end{figure}

Numerical orbitals of compact support were taken from the Siesta package 
\cite{siesta}. Their default support is 4\ldots 6 bohr which is
about two to three times smaller than the effective limit chosen for
the support of deMon2k's orbitals. Such a small support still allows to
reproduce basic features of electronic excitation spectra. A basis of larger support
would certainly improve the quality of spectra. In the LCAO technique,
the choice of basis is critical already for the ground state DFT calculation,
and one must check the basis again for the convergence of spectra
of excited states. Since this section is about testing our method of computing
spectra with our construction of $\chi_{0}$, we make no effort to investigate
errors related to the small support of this basis.

There is a substantial reduction in the number of dominant products when using
the default Siesta orbitals (of small spatial extent).
For instance, figure \ref{f:spectrum-benzene-siesta} shows a converged spectrum
of benzene in which the basis of dominant products is kept 7 times smaller than
original basis of localized products. To judge the completeness of the basis of
dominant products and the discretization errors, we provide a reference spectrum,
computed with the original products of molecular orbitals $\varphi _{E}(\bm{r})\varphi _{F}(\bm{r})$ 
\cite{Casida,Martin}. Due to unfavourable scaling behavior,
such a reference calculation is only possible for sufficiently small molecules like
benzene or naphthalene.

\section{Conclusions}

\label{s:5}

In this paper we have given an efficient construction of the Kohn--Sham response
function for molecular systems. To find $\chi_{0}$, we
made use of a previously found basis in the space of orbital products where
$\chi_{0}$ acts as a frequency dependent matrix. Our construction makes
extensive use of fast Fourier techniques and it requires $O(N^{2}N_{\omega })
$ \ operations for $N$ atoms on a lattice of $N_{\omega }$ frequencies. Two
approximations were made: a basis was chosen in the space of orbital
products with an error that vanishes exponentially in its size and the
electronic spectral densities were discretized.

We tested our construction directly on exact results for $\chi_{0}$ and by
calculating electronic excitation spectra. The comparison with the exact but
slow representation of $\chi_{0}$ showed good accuracy of our construction.
The excitation spectra from the Petersilka--Gossmann--Gross equations agreed
with those of Casida's equations. Moreover, an iterative Lanczos procedure
allowed us to maintain $O(N^{2}N_{\omega })$ scaling also for electronic
excitation spectra. In this approach, the CPU time grows less steeply than
in the solution of Casida's equations that requires $O(N^{3})$ operations.
The scaling of the Quantum Espresso method remains unpublished, but it is
likely to be $O(N^{2})$, according to one of its authors \cite{RalphGebauer-pc}.

Our construction of $\chi_{0}$ should have applications to excitons in
polymers or organic semiconductors where the Coulomb interaction is poorly
screened, and for implementing the GW approximation in molecular physics in
a straightforward way. It is also planned to use our algorithm for the
spectroscopy of surface adsorbed dyes.

\bigskip

\textbf{Acknowledgements}

It is a pleasure to thank James Talman (University of Western Ontario, London)
for contributing two crucial algorithms to this project, for making
unpublished computer codes of these algorithms available to us, and for many
fruitful discussions.

D.F. is grateful to Peter Fulde for extensive and continued support and for
inspiring visits at MPIPKS, Dresden that provided perspective for the
present work. Part of the collaboration with James Talman was done in the
pleasant environment of MPIPKS.

D.F. acknowledges the kind hospitality extended to him by Gianaurelio
Cuniberti and his collaborators at the Nanophysics Center of Dresden.

Both of us are indebted to Daniel Sanchez (DIPC, Donostia) for strong
support of this project and for advice and help on the Siesta code.

We also thank Andrei Postnikov (Paul Verlaine University, Metz) for useful advice.

Olivier Coulaud of the NOSSI project and (INRIA, Bordeaux) helped with the
Lanczos algorithm and by reading the manuscript. Our colleagues in this
project, Ross Brown and Isabelle Baraille (both at IPREM, Pau), Nguyen Ky
and Pierre Gay (both at DRIMM, Bordeaux), and Alain Marbeuf (CPMOH,
Bordeaux) have contributed with many useful discussions.

We thank Mark~E.~Casida and Bhaarathi Natarajan (Joseph Fourier University,
Grenoble) and their colleagues at Centro de Investigacion, Mexico for
letting us use their deMon2k code and for much help with it. Our special
thanks go to Mark~E.~Casida for his pertinent comments on our manuscript.

We also thank Stan van Gisbergen's (Vrije Universiteit, Amsterdam) for a
trial licence of ADF.

This work was financed by the French ANR project ``NOSSI'' (Nouveaux Outils
pour la Simulation de Solides et Interfaces). Financial support and
encouragement by ``Groupement de Recherche
GdR-DFT++'' is gratefully acknowledged.

\end{document}